# A 2D ferroelectric vortex lattice in twisted $BaTiO_3$ freestanding layers

G. Sánchez-Santolino[*,& 1], V. Rouco[*,& 1], S. Puebla[2,] H. Aramberri[3], V. Zamora[1], F. A. Cuellar[1], C. Munuera[2,4], F. Mompean[2,4], M. Garcia-Hernandez [2,4], A. Castellanos-Gomez[2,4], J. Íñiguez [3,5], C. Leon[1,4], J. Santamaria[1,4&].

[1] GFMC. Dept. Fisica de Materiales. Facultad de Fisica. Universidad Complutense. 28040 Madrid

[2] Instituto de Ciencia de Materiales de Madrid ICMM-CSIC 28049 Cantoblanco. Spain

[3] Materials Research and Technology Department, Luxembourg Institute of Science and Technology (LIST), Avenue des Hauts-Fourneaux 5, L-4362 Esch/Alzette, Luxembourg.

[4] Unidad Asociada UCM/CSIC, "Laboratorio de Heteroestructuras con aplicación en spintrónica"

[5] Department of Physics and Materials Science, University of Luxembourg, 41 Rue du Brill, L-4422 Belvaux, Luxembourg

[&]Corresponding authors: G. Sanchez-Santolino: gsanchezsantolino@ucm.es, V. Rouco: vrouco@ucm.es, SANTAMARIA Jacobo: jacsan@ucm.es

[*] Equal contributors

**The wealth of complex polar topologies [1- 6] recently found in nanoscale ferroelectrics result from a delicate balance between the materials' intrinsic tendency to develop a homogeneous polarization and the electric and mechanic boundary conditions imposed upon them. Ferroelectric–dielectric interfaces are model systems where polarization curling originates from open circuit-like electric boundary conditions, to avoid the build-up of polarization charges through the formation of flux-closure [7-10] domains that evolve into vortex-like structures at the nanoscale [11-13]. Interestingly, while ferroelectricity is known to couple strongly to strain (both homogeneous [14] and inhomogeneous [15, 16]), the effect of mechanical constraints [17] on thin film nanoscale ferroelectrics has been comparatively less explored because of the relative paucity of strain patterns that can be implemented experimentally. Here we show that the stacking of freestanding ferroelectric perovskite layers with controlled twist angles opens an unprecedented opportunity to tailor these topological nanostructures in a way determined by the lateral strain modulation associated to the twisting. Interestingly, we find that a peculiar pattern of polarization vortices and antivortices emerges from the flexoelectric coupling of polarization to strain gradients. This finding opens exciting**

**opportunities to create two-dimensional high density vortex crystals that would allow us to explore novel physical effects and functionalities.**

TEXT

The persistence of ferroelectricity at the nanoscale hinges on the compensation of the polarization bound charges and depolarizing fields building up at surfaces or interfaces. In ferroelectric films with metallic electrodes the depolarizing fields can be screened by (free) charge accumulation and by the formation of domains [18]. The situation is even more dramatic in nanoscale ferroelectric samples with dielectric boundaries (including vacuum or insulating non-polar surface layers) where the polarization can undergo a transition into vortex [11-13] or more complex [1-6] topological states, with rotational polar configurations persisting to small diameters where polarization departs from the high-symmetry directions favored by the lattice anisotropy [19].

Mechanical boundary conditions [17], as those imposed by interfacial strain, play a critical role in determining the final polarization state, as they may combine with electric boundary conditions in non-trivial ways. Importantly, the strong coupling of ferroelectricity to both homogeneous and inhomogeneous strain is at the origin of the effectiveness of mechanical boundary conditions in triggering unexpected effects, such as the enhanced ferroelectricity in epitaxially strained layers [14] or the polarization switching under the strain gradients created by an AFM tip pressing on the sample surface [20]. As it turns out, however, access to externally tunable strain patterns is in practice very limited.

In epitaxial thin films mechanical boundary conditions are to a large extent immovably and solely determined by the atom-on-atom replication of the structure of the substrate by the growing film. Hence, while the interface with the substrate is subject to in-plane strains imposed by the lattice mismatch, the sample surface is in a zero stress state, as there are no tractions acting on it. In epitaxial uniformly strained single-domain layers, internal elastic fields are homogeneous and rigidly imposed by these mixed boundary conditions. Inhomogeneous strain results typically from uncontrollable strain relaxation, misfit dislocations or ferroelastic domain formation [15]. The structural constraints imposed by epitaxy leave little or no room to modify mechanical boundary conditions. Moreover, *controllable* shear or inhomogeneous strain patterns are commonly out of reach. This is the reason why, although on general grounds exotic ferroelectric states can be expected to result from the manipulation of mechanical boundary conditions, this scenario remains mostly unexplored.

In this paper we demonstrate a new strategy to engineer mechanical boundary conditions based on the strain modulation induced at the interface between two twisted freestanding oxide layers. In layered materials such as graphite [21, 22] or transition metal dichalcogenides [23-25], twisted bilayers have led to the emergence of unexpected collective states [26, 27]. Interestingly, the weak van der Waals interlayer interaction in such twisted bilayers leads to inhomogeneous strain patterns with deformations up to 2% [28]. Extending the exploration to artificial twisted stacks of ionically bonded transition metal oxides, however, has been hampered by the difficulty to isolate these systems in freestanding form. The recent reports on the fabrication of freestanding single crystalline oxide thin films [29-31], which can be handled in a way similar to van der Waals 2D materials, open up the possibility of stacking freestanding layers with arbitrary twist angles [32, 33] and thus design completely novel strain patterns. To our knowledge, this approach to induce strain landscapes spatially varying at the nanoscale in complex oxides has not been reported so far. Here we show that the lateral strain modulation caused by the interface matching between two twisted freestanding ferroelectric $BaTiO_3$ layers sets a mechanical boundary condition not attainable by epitaxial strain, and to a large extent controllable by the relative rotation angle. The nanoscale modulated distribution of symmetric and antisymmetric shear strains yields a surprising rotational polarization texture with alternating clockwise and counterclockwise vortices and antivortices, whose distribution, spacing and size is controlled by the twist angle. First-principles simulations show that this complex configuration of highly localized symmetric and antisymmetric shears concomitant with the ferroelectric vortex 2D crystal constitutes in fact a stable equilibrium state. The coupling between shear strain gradients and complex polarization texture is discussed in terms of a direct flexoelectric effect.

15 nm thick $BaTiO_3$ (BTO) layers epitaxially grown on (001) $SrTiO_3$ (STO) substrates were delaminated to form twisted bilayer homojunctions with deterministic twist angles (see Figure 1a and Methods). Bilayers were transferred onto holey $Si_3N_4$ membranes. To study the structural properties of individual layers of the twisted bilayers we performed a depth sectioning high angle annular dark field (HAADF) scanning transmission electron microscopy (STEM) experiment (see Methods). Focusing on the entrance surface of the stack (defocus = 0 nm) we observe the typical structure of a $BaTiO_3$ perovskite which corresponds to the top layer. The Moiré contrast is revealed by changing the defocus to reach the interface of the twisted bilayer (defocus = - 15 nm), as shown in Fig. 1b. A further increasing defocus brings the bottom layer in focus, which appears rotated by the twist angle of the bilayer. Twisted ferroelectric bilayers exhibit characteristic Moiré features determined by the atomic coincidence pattern between the two layers (see Fig. 1c). The 10.4 º twist angle, determined from the fast Fourier transform (FFT) image, is homogeneous along the fabricated sample and close to the nominal 10º rotation of the films during the deterministic transfer process. The FFT shows the spots from both top and bottom twisted BTO layers; for clarity we will denote the directions corresponding to the twisted layer forming the Moiré pattern as $(100)^*$ and $(010)^*$. The Moiré pattern shows two distinct (plateau-like) features at the highly (atom-on-atom) coincidental regions of both layers, marked as AA and AB in Fig. 1b and on the rigid atomic model shown in Fig. 1d. Around AA sites there is AA stacking (Ba on Ba, Ti on Ti and O on O) between the top and bottom layers, while AB sites show an AB stacking (Ba on Ti and Ti on Ba) for the Ba and Ti cations of the twisted layers while preserving the AA stacking for the O anions. We studied Moiré structures formed in α = 3, 6, 10.4 and 50º twisted $BaTiO_3$ bilayers; in the main text we discuss

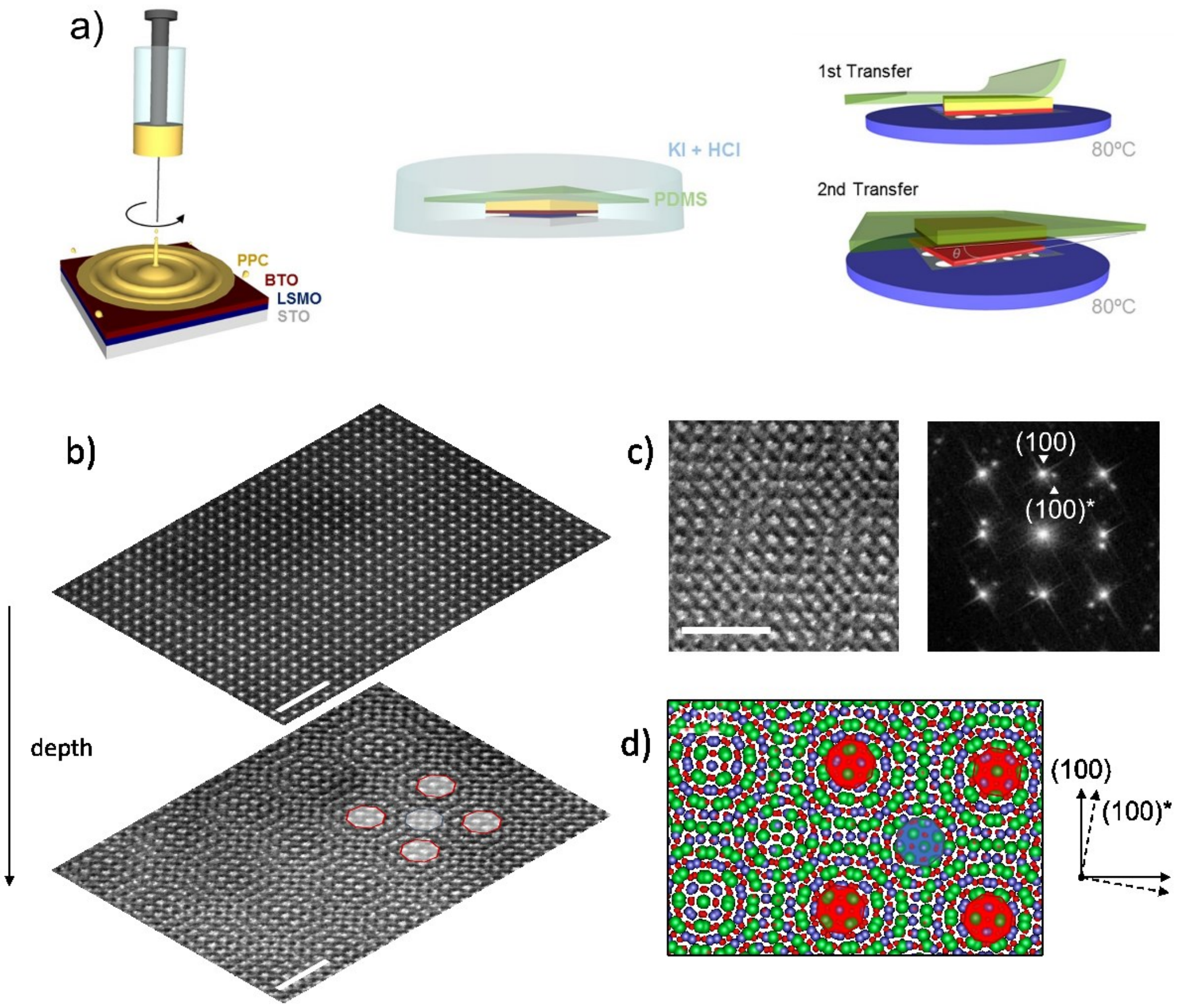

**Figure 1: Fabrication of twisted freestanding $BaTiO_3$ bilayers.** a) Schematic of the deterministic two steps transfer process. b) Depth sectioning STEM-HAADF experiment of a twisted $BaTiO_3$ bilayer stack focusing on the entrance surface of the stack (defocus = 0 nm) and the interface (defocus = -15 nm) c) Moiré structure formed at the interface between the $BaTiO_3$ layers. A network of two distinct AA and AB motives is formed due to the twist. The image on the right shows the Fast Fourier Transform noting the reflections of both the top and bottom freestanding $BaTiO_3$ films. The scale bar is 2 nm d) Model of the rigid atomic structure corresponding to two $BaTiO_3$ lattices with a twist angle of 10.4° showing Ba atoms in green, Ti in blue and O in red. AA and AB sites are marked in blue and red respectively.

data taken on the samples with 10.4° and 3° twist angles.

Intralayer strain was measured on the top layer using the entrance surface focused image (defocus = 0). The emergence of a strongly spatially varying strain landscape with the same periodicity as the Moiré lattice - and determined at the top layer - clearly demonstrates the strong interaction between the two twisted layers. The resultant strain map shows a periodically modulated pattern of symmetric shear strains ($\varepsilon_{xy} = \frac{1}{2}(\frac{\partial u_x}{\partial y} + \frac{\partial u_y}{\partial x})$) with alternating positive and negative shear strain cores (see Figs. 2b, e for bilayers twisted 10.4 and 3 degrees, respectively). Strain analysis included the antisymmetric components of the strain tensor ($\omega_{xy} = \frac{1}{2}(\frac{\partial u_x}{\partial y} - \frac{\partial u_y}{\partial x})$) associated to local rotations of the perovskite lattice (See Extended Figure S1 a and b). Control experiments on single BTO freestanding layers show a nearly homogeneous strain distribution, making it clear that the complex strain maps obtained in the bilayers originate at the stacking of the twisted layers. Notice that at the AA and AB sites there is maximal atom-on-atom coincidence between the twisted layers; we find very small shear strains in those areas. In between the AA and AB sites of the Moiré pattern we find regions of maximum strain, named S-sites hereafter, with nearly homogeneous positive and negative shears. The shear strain modulation shows the same periodicity as the Moiré pattern, indicating that the strain results in fact from a displacement field reconstruction in the top layer induced by the matching at the interface. An important remark is that such a periodical shear strain landscape is unique, as it cannot be attained, to the best of our knowledge, either by epitaxial strain or by any pattern of externally applied stresses.

In order to investigate how the strain modulation observed on the top layer of the twisted BTO bilayers affects the ferroelectric polarization we have measured the off-centering of the B-site cations in the individual unit cells (relative displacement of the B-site Ti cation from the centrosymmetric position, determined with the A-site Ba cations within the same unit cell). Twisted bilayers showed net in-plane polarization in the [1,1] direction of the perovskite lattice (in the pseudo-cubic reference) with a superimposed polar texture. Polar displacements were obtained to be in the range of 0.15 to 0.20 Å, which is consistent with what is found in bulk $BaTiO_3$. In $BaTiO_3$, the magnitude of the spontaneous polarization is known to be approximately constant regardless of the orientation it may present in the different ferroelectric phases this compound can adopt [34, 35]. This suggests that the polarization of our layers must be largely confined to the plane, the (not measured) vertical component being very small if present at all. Further, this in-plane polarization is most likely a consequence of the stacking in our materials. The interfaces between layers will inevitably result in a discontinuity of the vertical component of the electric displacement vector, with the concomitant occurrence of depolarizing fields. In such conditions, an in-plane polarization is energetically favored, as it does not suffer from any electrostatic penalty. The situation can be compared to that of layered perovskite crystals (e.g., Ruddlesden-Popper phases), which typically present spontaneous polarizations perpendicular to the stacking direction [36-38]. Note also that our free-standing layers are not subject to any global mechanical constraints that might drive the occurrence of an in-plane polarization, further supporting its electrostatic origin.

The net polarization pointing along the $[1,1]$ in-plane direction indicates that our layers present the orthorhombic ferroelectric phase that also occurs in the bulk material. $BaTiO_3$ crystals present a tetragonal structure at room temperature, which would suggest a polarization along $[1,0]$ or $[0,1]$ in our layers; yet, single BTO freestanding layers (see Extended Figure S2) showed averaged polarization vectors in the [1,1] direction. Given the proximity to the bulk tetragonal to orthorhombic transition (which occurs at 278 K) and the fact that these two phases have very similar free energies at ambient conditions, the observed orthorhombic-like state seems perfectly acceptable, as it may be stabilized by any of many factors distinguishing our $BaTiO_3$ layers from the bulk compound.

The complex polar texture of our layers can be better assessed by subtracting the average polarization value in the image (See Extended Figure 3 for a comparison). The polarization maps in Figs. 2c, f show a continuous curling of the polar displacements, forming a periodic network of non-trivial topological structures with alternating polarization vortices (AA and AB sites) and antivortices (S sites of the Moiré pattern). We can describe the topological structure in terms of a non-zero toroidal moment [11- 13] parallel to the z direction defined as $Q = \frac{1}{2N}\sum_1^N r_i x P_i$, where $P_i$ is the local dipole moment located at $r_i$

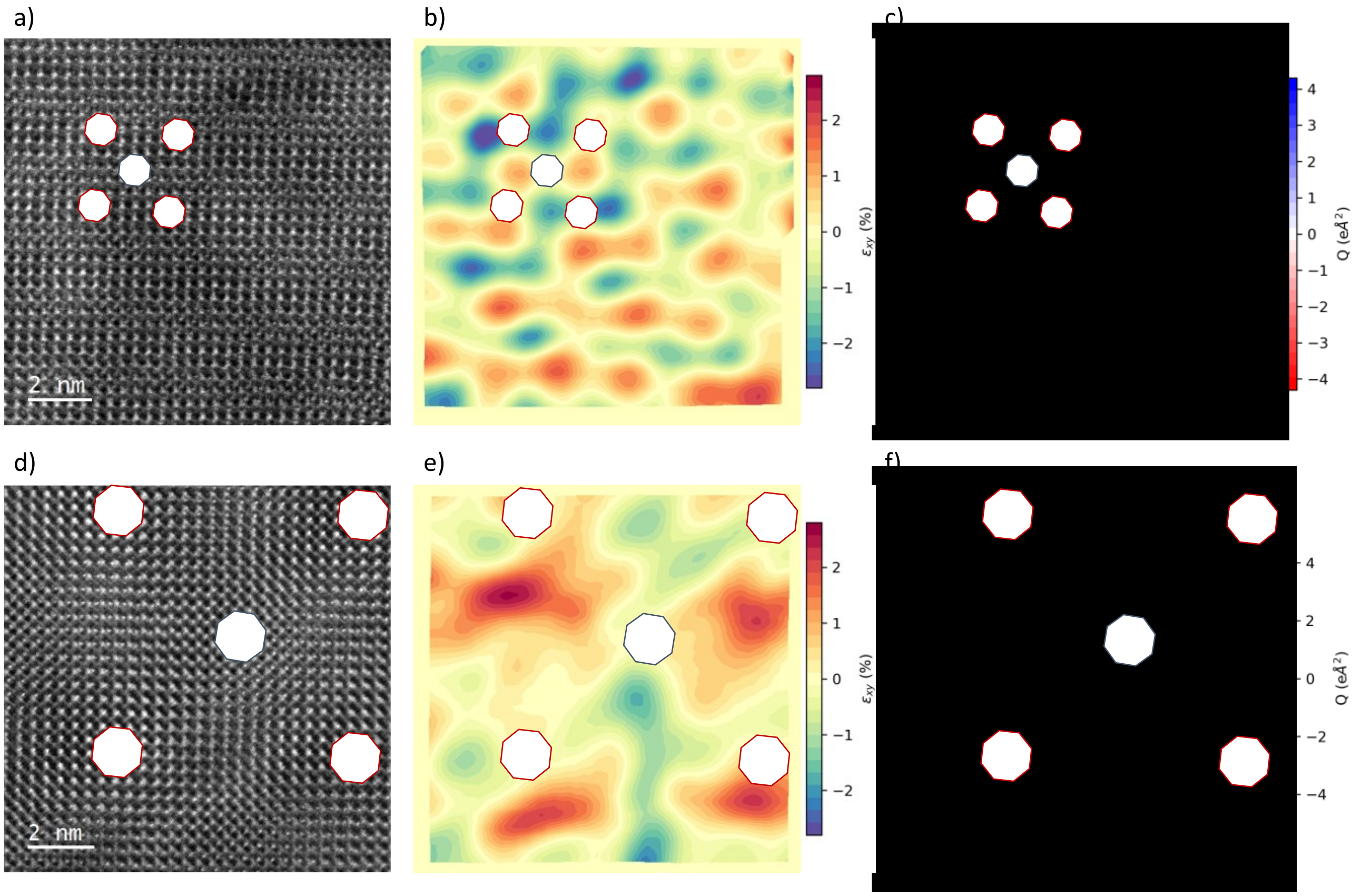

**Figure 2: Strain and polarization modulations at twisted $BaTiO_3$ bilayers.** a) STEM-HAADF image of a 10.4° twisted $BaTiO_3$ bilayer stack focusing on the interface of the bilayer (defocus = -15 nm). b) Shear strain ($\varepsilon_{xy}$ component of the lattice strain tensor) depicting a periodic strain modulation at the top $BaTiO_3$ layer. c) Ti displacement map (black arrows) measured on the top $BaTiO_3$ layer corresponding to the same area superimposed to the toroidal moment (Q) of the ferroelectric polarization showing a network of clockwise (red) and counterclockwise (blue) vortices. Ti displacements are amplified by a factor of 20 for clarity. d), e), f) show the same analysis for a 3° twisted $BaTiO_3$ bilayer. Red and blue octagons in all panels indicate sites with AA (AA-sites) and AB (AB-sites) stacking respectively. The averaged polarization (modulus) is approximately 20 $\mu C\ cm^{-2}$, close to the bulk $BaTiO_3$ value.

and N is the number of dipoles (cells).

The toroidal moment alternates sign periodically in diagonal directions of the Moiré pattern (see Figures 2c and 2f) in a way determined by a periodic array of alternating clockwise and counterclockwise vortices in AA and AB sites, respectively. Ferroelectric vortices are topological objects characterized by a winding number $n$ =+1 (see Supplementary Note 1) regardless of their polarity (clockwise or counterclockwise). Values of the toroidal moment at the vortex sites depend on the size of the vortex and on the ferroelectric displacements (dipole moment). Interestingly, we obtain values similar to those reported for flat epitaxial BTO nanoparticles [13]. In the Moiré pattern, vortices alternate with antivortices (sitting at S-sites), which are topological structures with $n$ =-1 winding number and zero toroidal moment. Crucially, there is a close correspondence between the vortex lattice with the distribution of shear strains underlying the Moiré pattern. Clockwise or counterclockwise vortices are located at AA- and AB-sites with nearly zero shear strain (albeit maximal rotational strain). On the other hand, antivortices sit at the S sites with maximal shear strain (but nearly zero rotational strain).

To get further confirmation of this topological polar pattern, we resort to density-functional theory, considering simplified (computationally tractable) simulated systems that are nevertheless relevant to our problem. More precisely, we work with a periodically repeated supercell composed of 6x6x1 elemental $BaTiO_3$ units and consider an initial configuration that mimics the inhomogeneous polarization pattern measured in the 10° twisted layers. We then run a structural relaxation where all variables can evolve to minimize the energy of the system. We obtain, as a stable solution, the polarization and strain maps shown in Fig. 3, in qualitative agreement with the experimental results of Fig. 2 and thus confirming the connection between the observed strain and dipole modulations. According to our simulations, this topological state is 9 meV per formula unit above the homogeneous orthorhombic phase with polarization along the [1,1] diagonal. This relatively small difference is in fact an upper bound (see Methods) for the energy cost of deforming the trivial homogeneous state to acquire the topological features of Figs. 2 and 3. Hence, our calculations support the notion that interlayer interactions may suffice to induce the experimentally observed strain and dipole patterns.

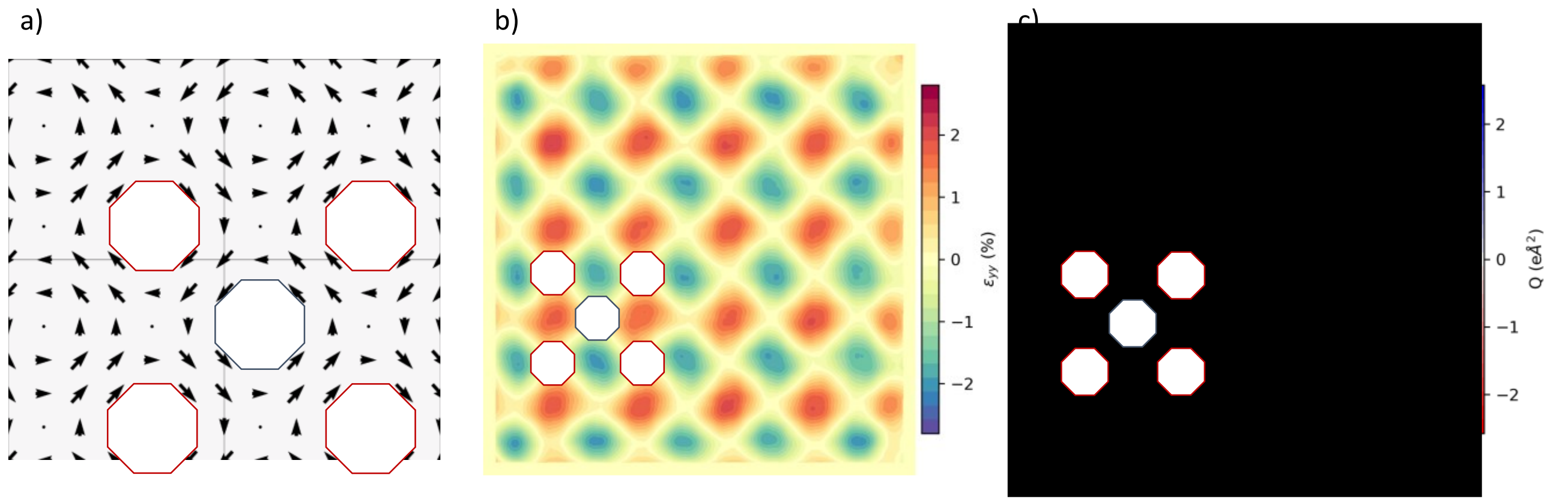

**Figure 3: Density functional theory model of a ferroelectric vortex lattice in BaTiO₃.** a) DFT calculated model. b) Shear strain ($\varepsilon_{xy}$ component of the lattice strain tensor) obtained from the DFT model. c) Ti displacement map (black arrows) superimposed to the toroidal moment (Q) of the ferroelectric polarization obtained from the DFT model. Ti displacements are amplified by a factor of 40 for clarity. Red and blue marks in all panels indicate the AA and AB stacking regions, respectively.

Let us finally tackle this critical question: what causes the peculiar inhomogeneous polarization textures in our layers? These complex quasi-periodic orders are controlled by the twist angle, which indicates they are the result of interlayer interactions. Further, it is apparent that the vortex- and antivortex-like dipole arrangements in Figs. 2c and 2f are correlated with the measured strain patterns of Figs. 2b and 2e. This suggests that, to understand these polar textures, it is reasonable to ignore the microscopic details of the couplings across the twisted interface and, instead, focus on how the observed elastic modulation affects the polarization. Indeed, ferroelectric perovskites like $BaTiO_3$ present strong electromechanical couplings that are potential candidates to explain our observations.

Let us begin by considering the simplest strain-polarization couplings. From well-established models of ferroelectric perovskites like $BaTiO_3$ [39], we know that a shear strain $\varepsilon_{xy} > 0$ typically favors a polarization oriented along the $[1,1]$ in-plane diagonal, while $\varepsilon_{xy} < 0$ leads to polarizations along $[1,-1]$; hence, we can expect $\delta P_x \delta P_y \propto \varepsilon_{xy}$, where by $(\delta P_x, \delta P_y)$ we refer to the inhomogeneous part of the measured polarization, as shown in Figs. 2c and 2f (and also Fig. 3c). However, it is clear from our results that this relationship does not hold for the measured strains (Figs. 2b and 2e, and Fig. 3b) and inhomogeneous polarizations (Figs. 2c and 2f, and also Fig. 3c), as one can e.g. find regions with $\varepsilon_{xy} > 0$ and an either positive or negative $\delta P_x \delta P_y$ product. A strong piezoelectric effect would also lead to $\delta P_x \delta P_y \propto \varepsilon_{xy}$, and is not supported by our observations either. Hence, these are not the dominant couplings in our samples.

Next, we note that our measured strain maps feature large strain gradients with maximum values reaching $\pm 4 \times 10^7\ m^{-1}$, which we explicitly show in Figure 4. By virtue of the direct flexoelectric coupling [38], such gradients should yield a polarization change, the expected dominant effects being

$$\delta P_x \approx \mu_{xyxy}^{\text{eff}} \frac{\partial \epsilon_{xy}}{\partial y} \qquad (1)$$

and

$$\delta P_y \approx \mu_{xyxy}^{\text{eff}} \frac{\partial \epsilon_{xy}}{\partial x} \qquad (2)$$

where $\mu_{xyxy}^{\text{eff}}$ is the effective flexoelectric coefficient active in our samples. Notably, from the measured strain gradients (Figs. 2b and 2e, and Fig. 3b) and inhomogeneous polarization (Figs. 2c and 2f, and also Fig. 3c), we do see direct support for this coupling in our results. In fact, we find that the regions with $\frac{\partial \epsilon_{xy}}{\partial x} > 0$, shown as red vertical fringes in Figure 4, feature positive $\delta P_y > 0$; conversely, the regions with $\frac{\partial \epsilon_{xy}}{\partial x} < 0$, shown as blue vertical fringes in Figure 4, show $\delta P_y < 0$. A similar relation holds for the $\frac{\partial \epsilon_{xy}}{\partial y}$ gradients and the $\delta P_x$ component of the polarization.

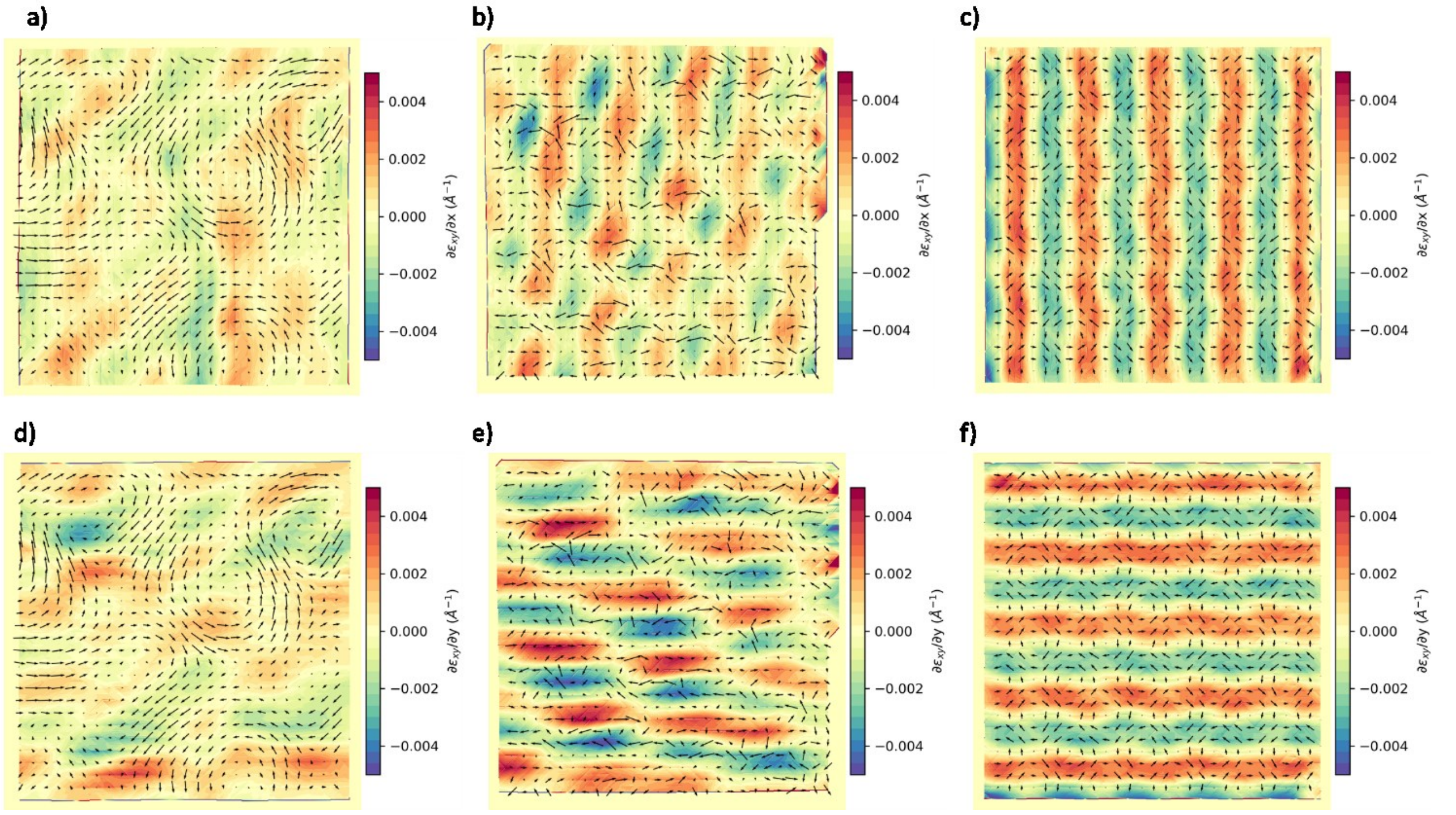

**Figure 4: Shear strain gradients of twisted $BaTiO_3$ bilayers.** Derivative of the shear strain along the x axis of a 3° twisted $BaTiO_3$ bilayer (a) and a 10,4° twisted $BaTiO_3$ bilayer (b) and a DFT calculated model corresponding to 10° twisted layers (c). Derivative of the shear strain along the y axis of a 3° twisted $BaTiO_3$ bilayer (d) and a 10.4° twisted $BaTiO_3$ bilayer (e) and a DFT calculated model corresponding to 10° twisted layers (f). Ti displacement map (black arrows) are superimposed to all images. Ti displacements are amplified for clarity by a factor of 20 in (a, b, d, e) and 40 in (c, f).

In fact, the relationship between strain and polarization patterns can be captured in a simple geometric manner. As shown in Fig. 5, the symmetry breaking caused by the shear (and rotational) strain modulation readily leads to the observed arrangement of polar vortices and antivortices. A local shear strain $\epsilon_{xy} \neq 0$ breaks the square symmetry of the cells of Fig. 5, yielding two large-angle corners and two small-angle corners. In the figure, arrows (flexoelectric polarizations) are drawn assuming that the cations displace towards the small-angle corners, which naturally yields an antivortex-like dipole arrangement with zero curl of the polarization field centered at the cells with $\epsilon_{xy} \neq 0$. Correspondingly, polarization vortices (non-zero curl) form around the cells with $\epsilon_{xy} = 0$.

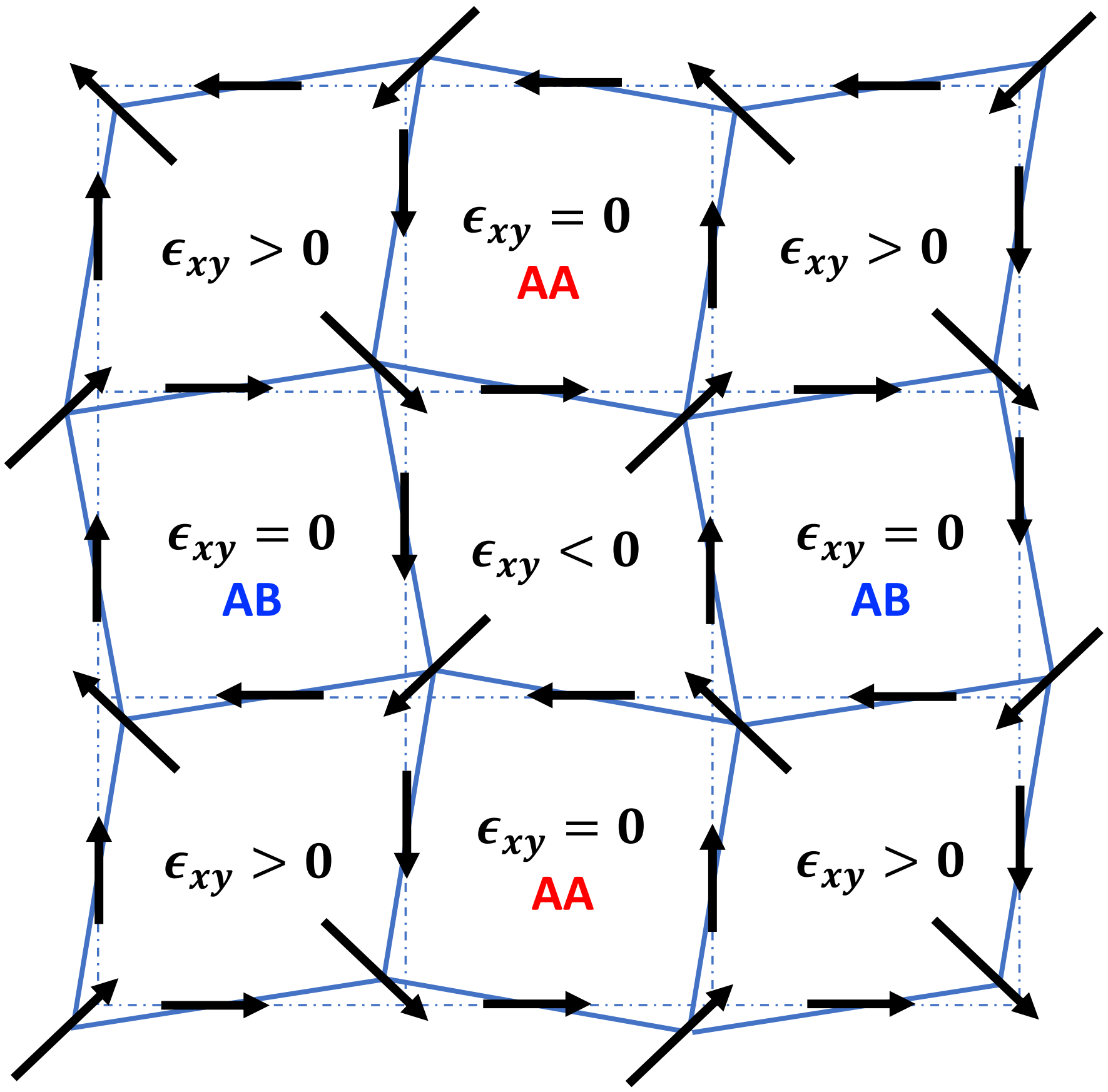

**Figure 5: Pictorial view of the flexoelectric couplings.** Sketch of the $BaTiO_3$ layer, showing regions of approximately constant shear strain as cells of a periodic lattice. We indicate the analogues of the AA and AB sites discussed in the text. The black arrows stand for the polarization induced by the flexoelectric effect; these arrows are consistent with Eqs. (1) and (2) for $\mu^{\text{eff}}_{xyxy} > 0$, and they present the vortices and antivortices observed experimentally. Note that the flexoelectric polarization can be intuitively understood from the symmetry breaking caused by the strain modulation. For example, at any given lattice point (shared by four cells, with four associated cell angles), we always find an arrow pointing towards the cell with the smallest ($< 90°$) angle.

Our quantitative measurements allow us to compute strain gradients and polarization modulations and, thus, estimate the effective flexoelectric coefficient $\mu^{\text{eff}}_{xyxy}$. We approximately have

$$\mu_{xyxy}^{\text{eff}} \approx \delta P_x \left(\frac{\Delta\epsilon_{xy}}{\Delta y}\right)^{-1} \approx \frac{20\ \mu C\ cm^{-2}}{4 \times 10^7\ m^{-1}} \approx 5\ nC\ m^{-1},$$

which is significantly smaller than typical experimental results for bulk $BaTiO_3$ at room temperature (values between 0.15 and 3.3 $\mu C\ m^{-1}$ have been reported [40-42]). A variety of reasons may explain this difference. For example, our constrained $BaTiO_3$ layers might be electrically stiffer than the bulk material and thus present a smaller flexoelectric response. (The magnitude of the flexoelectric coupling is known to be proportional to the magnitude of the dielectric response [17].) Probably most critical: a linear approximation as that in Eqs. (1) and (2) may be inadequate to explain and quantify the effect of the giant strain gradients in our samples. (To determine flexoelectric coefficients experimentally, the considered strain gradients are intentionally small, of the order of 1 $m^{-1}$ [43]. Strain gradients of the order of $8 \times 10^5\ m^{-1}$ – as those associated to ferroelastic domains [15] – are considered to be very large. The gradients in our samples are even larger, by almost two orders of magnitude.) Additionally, we may have differences coming from surface contributions to the flexoelectric effect [44]; such surface effects might be present in bulk measurements but seem unlikely to play a role in our case, since the relevant shears and gradients do not involve any component normal to the layers. Having said this, let us also note that there are theoretical predictions yielding $\mu_{xyxy}$ values around 0.08 $nC\ m^{-1}$ for $BaTiO_3$ [45], i.e., a smaller effect than the one we estimate. Shedding further light into these issues would be a great challenge, for both experiment and theory, and falls beyond the scope of the present work.

It is also interesting to note that the second derivatives of the shear strain can be used to compute the expected curl of the polarization vector. In fact, from the flexoelectric coupling between strain gradients and polarization (Eqs. (1) and (2)), the following relation holds:

$$\frac{\partial P_x}{\partial y} - \frac{\partial P_y}{\partial x} = \mu_{xyxy}^{\text{eff}} \left(\frac{\partial^2 \epsilon_{xy}}{\partial y^2} - \frac{\partial^2 \epsilon_{xy}}{\partial x^2}\right),$$

closely captured by experimental results (see Extended Figures S4 showing the second derivatives of the strain gradient and S5 showing the curl of the polarization).

In summary, we have found that the stacking of twisted free standing ferroelectric layers features a non-trivial ferroelectric texture driven by the mechanical boundary conditions imposed by the interface of twisted freestanding layers. The ferroelectric topology consists of a 2D vortex crystal with a lattice periodicity determined by the twisting angle. This opens the door to new design possibilities enabled by the unique modulations that are possible in Moiré bilayers. The highly correlated topological pattern with vortices and antivortices is reminiscent of the square lattice of merons, objects with n= ½ topological number only existing in lattices, observed in chiral magnets with magnetic anisotropy [46, 47]. At variance with previous ferroelectric textures found in ferroelectric films confined in the growth direction, our polar landscape is 2D and highly tunable by controlling the twisting angle of the bilayer and is, thus, more amenable for applications in high density ferroelectric memories.

**Materials and Methods:**

Freestanding perovskite films fabrication:

15 nm thick BTO layers were grown onto LSMO buffered (100) $SrTiO_3$ (STO) substrates via pure oxygen sputtering technique at high pressures (3.2 mbar) [48]. This technique produces highly epitaxial growth with sharp interfaces and negligible stoichiometry deviations (see Extended Figure S6). The LSMO acts as a sacrificial layer that allows the release of the BTO layer upon immersion in a selective KI+HCl etchant [49]. Prior to immersion, a polypropylene carbonate (PPC, Sigma Aldrich) film was spin-coated onto the strained heterostructure and adhered to a commercial polydimethylsiloxane (PDMS, Gel-Film WF 4x 6.0 mil by Gel-Pak®) support. This allows the release and transfer of the entire BTO freestanding layer onto holey $Si_3N_4$ membrane grids for STEM observation. After the first and prior to the second BTO transfer, the membranes are dipped in acetone and isopropyl alcohol to remove the remained PPC and clean the final interface. The second BTO layer is transferred onto the first one with a twisted angle which is deterministically controlled by using the edges of the two BTO layers (with defined crystallographic orientation imposed by the (100) STO substrate) as a reference. See sketch in Fig. 1a. After the second transfer the surface of the final twisted heterostructure is cleaned as described above.

Scanning transmission electron microscopy (STEM):

STEM characterization was carried out using a JEOL JEM-ARM 200cF aberration corrected electron microscope equipped with a cold field emission gun and a Gatan Quantum spectrometer, operated at 200 kV. Depth sectioning HAADF-STEM was performed by acquiring atomic-resolution HAADF-STEM images as a function of defocus [50, 51], allowing us to probe different depths of the sample and discriminate between the top and bottom layers of the stack. HAADF-STEM images were acquired using a 30 mrad probe forming aperture semiangle and a HAADF detector collection semiangle of 70-200 mrad.

Determination of polarization and strain.

To determine the ferroelectric polarization, the atomic positions of both A-site Ba and B-site Ti cations were measured on STEM-HAADF images of 3° and 10,4° twisted $BaTiO_3$ bilayer stacks acquired focusing on the entrance surface of the stack (defocus = 0 nm). In order to precisely determine the atomic positions, we performed a two-dimensional gaussian fitting using Atomap [52] Polarization was calculated from the off centering of the B-site Ti cations in the individual unit cells (relative displacement of the B-site Ti cation from the centrosymmetric position, determined with the A-site Ba cations within the same unit cell) [53].

Strain analysis was performed using the Peak Pairs Analysis (PPA) software package (HREM Research) for Digital Micrograph [51]. The analysis was performed on STEM-HAADF images of 3° and 10,4° twisted $BaTiO_3$ bilayer stacks acquired focusing on the entrance surface of the stack (defocus = 0 nm). In order to improve the precision of the analysis the scanning direction was rotated off the crystallographic axes of $BaTiO_3$. For the analysis we perform a Bragg filter selecting the two main reflections along the (100) and (010) directions as base vectors for the analysis. The peak positions are then determined on the filtered image and the relative displacements fields ($u_x$, $u_y$) of the measured lattice with respect to a reference lattice area are calculated. In this case we have use the whole image as reference. Finally, the components of the strain tensor are calculated from the displacement fields as: $\varepsilon_{xx} = \frac{\partial u_x}{\partial x}$, $\varepsilon_{yy} = \frac{\partial u_y}{\partial y}$, $\varepsilon_{xy} = \frac{1}{2}\left(\frac{\partial u_x}{\partial y} + \frac{\partial u_y}{\partial x}\right)$ and $\omega_{xy} = \frac{1}{2}\left(\frac{\partial u_y}{\partial x} - \frac{\partial u_x}{\partial y}\right)$.

First-principles calculations:

We performed density functional theory (DFT) calculations as implemented in the Vienna Ab initio Simulation Package (VASP) [54, 55] . We used the Perdew-Burke-Ernzerhof formulation for solids (PBEsol) [56] implementation of the generalized gradient approximation for the exchange-correlation functional. The atomic cores are treated within the projector-augmented wave approach [57], considering the following states explicitly: 5s, 5p and 6s for Ba; 3p, 4s and 3d for Ti; and 2s and 2p for O. We employed a 500 eV energy cut-off for the plane-wave basis set. The simulation cells comprised 6x6x1 perovskite unit cells and were computed using a 1x1x4 Monkhorst-Pack [58] k-point grid. The structures were fully relaxed until residual forces fell below 0.01 eV/Å and residual stresses fell below 0.001 GPa.

Let us stress that our DFT simulations correspond to the limit of very low temperature (formally, 0 K). Thus, the computed energy differences – i.e., the 9 meV per formula unit separating the monodomain ferroelectric state from the vortex-antivortex structure – can be taken as an upper bound for the relevant free energy difference at room temperature. (In essence, the calculated energy difference comes from the ferroelectric domain walls – whose energy is known to decrease upon heating – and the inhomogeneous strain modulation – which is imposed by the inter-layer couplings.) Note also that in our simulations we treat the monodomain and vortex-antivortex configurations as two separate cases, while in experiment the topological features are a relatively small modulation of the homogeneous state. For this reason too, the computed energy difference is an upper bound for the actual energy cost of inducing (relatively small) topological features superimposed to the homogeneous state. All in all, our DFT results strongly suggest that the experimentally observed topological structure is easily accessible and physically sound. Finally, let us remark that simulating directly the perturbed homogeneous state would require DFT relaxations constrained to respect the experimentally observed inhomogeneous strain pattern; such calculations would involve several non-trivial assumptions and technical complications, and we did not pursue them here.

**Acknowledgments**

Authors acknowledge received funding from the project To2Dox of FlagERA ERA-NET Cofund in Quantum Technologies implemented within the European Union's Horizon 2020 Program. Work (JS, CL, FM , M G-H) supported by Spanish AEI through grants, PID2020-118078RB-I00 and by Regional Government of Madrid CAM through SINERGICO project Y2020/NMT-6661 CAIRO-CM. G.S.-S. acknowledges financial support from Spanish MCI Grant Nos. RTI2018-099054-J-I00 (MCI/AEI/FEDER, UE) and IJC2018-038164-I. Work (VR) supported by the Madrid Government (Comunidad de Madrid-Spain) under the Multiannual Agreement with Universidad Complutense de Madrid in the line Research Incentive for Young PhDs, in the context of the V PRICIT (Regional Programme of Research and Technological Innovation. European Union's Horizon 2020 research and innovation program (Grant Agreement No. 755655 ERC-StG 2017 project 2D-TOPSENSE, Grant Agreement No. 785219 Graphene Core2-Graphene-based disruptive technologies and Grant agreement No. 881603 Graphene Core3-Graphene-based disruptive technologies), the EU FLAG-ERA project To2Dox (JTC-2019-009), the Comunidad de Madrid through the CAIRO-CM project (Y2020/NMT-6661) and the Spanish Ministry of Science and Innovation (grant PID2020-118078RB-I00 and fellowship PRE2018-084818).

Electron microscopy observations were carried out at the Centro Nacional de Microscopia Electrónica, CNME-UCM. Work at LIST was supported by the Luxembourg National Research Fund through grant FNR/C18/MS/12705883/REFOX.

**Data availability statement.** The data used in this paper are available from the authors upon reasonable request

**Authors contributions.** VR, VZ, SP, FAC prepared the samples with help and guidance of CM, FM, MGH and ACG. GSS did the electron microscopy. GSS, VR, VZ, CL, JS analyzed the electron microscopy data. HA and JI did the theory analysis. GSS, VR, HA, JI, CL and JS wrote the manuscript with inputs and help of all authors.

**Competing interests:** The authors declare no competing interests.